\pdfoutput=1

\documentclass[12pt]{iopart}

\usepackage{times}
\usepackage{amsfonts}
\usepackage{amssymb}
\usepackage{cancel}
\usepackage{graphicx}
\usepackage{mathcomp}
\usepackage{multicol}
\usepackage[utf8]{inputenc}
\usepackage{color}
\usepackage{shadow}
\usepackage{times}
\usepackage[superscript,nospace]{cite}
\usepackage{booktabs}

\begin{document}

\newcommand{\rB}{{\rm B}}
\newcommand{\rF}{{\rm F}}
\newcommand{\rDD}{{\rm 3D}}
\newcommand{\rED}{{\rm 1D}}

\newcommand{\cH}{\mathcal{H}}
\newcommand{\cO}{\mathcal{O}}
\newcommand{\cI}{\mathcal{I}}
\newcommand{\cZ}{\mathcal{Z}}

\newcommand{\tcH}{\tilde{\cH}}

\newcommand{\hP}{\hat{\Psi}}
\newcommand{\hPd}{\hat{\Psi}^\dagger}
\newcommand{\hPdp}{\left.\hat{\Psi}^\dagger\right.'}
\newcommand{\hPe}{\hat{P}}
\newcommand{\hH}{\hat{H}}
\newcommand{\hM}{\hat{M}}
\newcommand{\hA}{\hat{A}}
\newcommand{\hAd}{\hat{A}^\dagger}
\newcommand{\hB}{\hat{B}}
\newcommand{\hX}{\hat{X}}
\newcommand{\hL}{\hat{L}}
\newcommand{\ha}{\hat{a}}
\newcommand{\had}{\hat{a}^\dagger}
\newcommand{\hsig}{\hat{\sigma}}
\newcommand{\hc}{\hat{c}}
\newcommand{\hcd}{\hat{c}^\dagger}
\newcommand{\hr}{\hat{\rho}}
\newcommand{\hN}{\hat{N}}
\newcommand{\hn}{\hat{n}}
\newcommand{\hU}{\hat{U}}
\newcommand{\hV}{\hat{V}}
\newcommand{\hE}{\hat{E}}
\newcommand{\hS}{\hat{S}}
\newcommand{\hI}{\hat{I}}
\newcommand{\hO}{\hat{O}}
\newcommand{\hOd}{\hat{O}^\dagger}
\newcommand{\ho}{\hat{o}}
\newcommand{\hod}{\hat{o}^\dagger}
\newcommand{\hId}{\hat{I}^\dagger}

\newcommand{\tO}{\tilde{O}}
\newcommand{\tH}{\tilde{H}}
\newcommand{\tT}{\tilde{T}}
\newcommand{\tsig}{\tilde{\sigma}}
\newcommand{\tSig}{\tilde{\Sigma}}

\newcommand{\hcI}{\hat{\cI}}

\newcommand{\bv}{{\bf v}}
\newcommand{\bt}{{\bf t}}
\newcommand{\be}{{\bf e}}
\newcommand{\bF}{{\bf F}}
\newcommand{\bnl}{{\bf nl}}

\newcommand{\scl}{\hspace{-0.6cm}}
\newcommand{\scls}{\hspace{-0.3cm}}
\newcommand{\ko}[2]{\left[ #1, #2 \right]}
\newcommand{\ako}[2]{\left\{ #1, #2 \right\}}
\newcommand{\expv}[1]{\langle #1 \rangle}
\newcommand{\Bigket}[1]{\Big\vert#1\Big\rangle}
\newcommand{\ket}[1]{\vert#1\rangle}
\newcommand{\bra}[1]{\langle#1\vert}

\newcommand{\hOg}{\hat{O}_{g}}
\newcommand{\eps}{\varepsilon}

\newcommand{\nl}{\overleftarrow{n}}
\newcommand{\nr}{\overrightarrow{n}}
\newcommand{\Pbl}{\overleftarrow{P}}
\newcommand{\Pbr}{\overrightarrow{P}}

\newcommand{\pnp}{\phi(0^+)}
\newcommand{\pnpp}{\phi'(0^+)}
\newcommand{\pnppp}{\phi''(0^+)}

\newcommand{\Ot}{\tilde{\Omega}}

\newcommand{\binom}[2]{\Big(\!\!\begin{array}{c}#1\\#2\end{array}\!\!\Big)}


\title[Particle number conservation in matrix product operators]{Particle number conservation in quantum many-body simulations with matrix product operators}

\author{Dominik Muth$^{1, 2}$}
\address{$^1$ Fachbereich Physik und Forschungszentrum OPTIMAS, Technische Universit\"at Kaiserslautern, Erwin-Schr\"odinger-Str. 46, D-67663 Kaiserslautern}
\address{$^2$ Graduiertenschule Materials Science in Mainz, Technische Universit\"at Kaiserslautern, Erwin-Schr\"odinger-Str. 46, D-67663 Kaiserslautern}
\ead{muth@physik.uni-kl.de}

\begin{abstract}
Incorporating conservation laws explicitly into matrix product states (MPS) has proven to make numerical simulations of quantum many-body systems much less resources consuming. We will discuss here, to what extent this concept can be used in simulation where the dynamically evolving entities are matrix product operators (MPO). Quite counter-intuitively the expectation of gaining in speed by sacrificing information about all but a single symmetry sector is not in all cases fulfilled. It turns out that in this case often the entanglement imposed by the global constraint of fixed particle number is the limiting factor.
\end{abstract}

\pacs{	02.70.-c, 
05.10.Cc, 
05.30.-d, 
05.50.+q 
}

 
\date{\today}
 
\maketitle


\section{Introduction}
\label{sec:int}

Variational MPS methods have been used for more than half a century \cite{Kramers1941b} to describe the transfer matrix of two-dimensional classical models in statistical mechanics, which are equivalent to one-dimensional quantum systems. For references see, e.g., the work of Baxter \cite{Baxter1978} and references therein. Later on the density-matrix renormalisation group (DMRG) method \cite{White1992} has been developed independently and proved very successful in describing low-energy eigenstates of one-dimensional quantum lattice systems which are typically only moderately entangled. In the last decade DMRG has been extended to real-time evolution \cite{Vidal2003, Vidal2004, Daley2004, White2004} (t-DMRG). These and various other extensions all rely on the MPS framework to capture the relevant part of the Hilbert space in terms of the largest singular values \cite{McCulloch2007, Schollwock2011}.

Conservation laws resulting from global symmetries can be taken into account explicitly in the construction of MPSs \cite{Schollwoeck2005, McCulloch2007, Singh2010, Bauer2011}. This reduces the number of degrees of freedom such that approximations with higher matrix dimensions can be calculated using the same amount of computation time and memory. In addition arithmetical errors of the type that would lead out of the symmetry sector of the initial state are impossible.

Implementing abelian symmetries is particularly easy \cite{Schollwoeck2005}. When calculating low lying eigenstates with a given accuracy, the gain in CPU time and memory is typically of an order of magnitude or more. In dynamical simulations, abelian symmetries allow for calculations on longer timescales\cite{Daley2005a}.

Particle number conservation, which results from a global $\textrm{U}(1)$ symmetry, is present in many non-relativistic model systems and implemented in MPS algorithms routinely. Its explicit implementation is necessary if one wants to calculate ground state\cite{Muth2010} or dynamical\cite{Muth2010b,Muth2010a} properties in the low filling limit, where the average number of particles per lattice site is small compared to $1$, as it results e.g. from the discretization of a continuous model \cite{Muth2010}.

In DMRG like dynamical simulations matrix product \emph{operators} naturally arise either as density-operators at non-zero-temperature \cite{Verstraete2004a, Zwolak2004} or in open systems \cite{Verstraete2004a, Zwolak2004, Hartmann2009, Prosen2009} or as general operators in the Heisenberg picture \cite{Prosen2007a}. (The Hamiltonian itself can also be conveniently expressed \cite{McCulloch2007} as an MPO of small dimension in the case of short-range interactions, which gives rise to elegant formulations of the algorithm \cite{Schollwock2011}.) In this paper we will focus on operators in the Heisenberg picture. Most of the results are however equally valid in the context of finite temperature calculations.

\begin{figure*}[htb!]
\centering
\includegraphics[width=.6\textwidth]{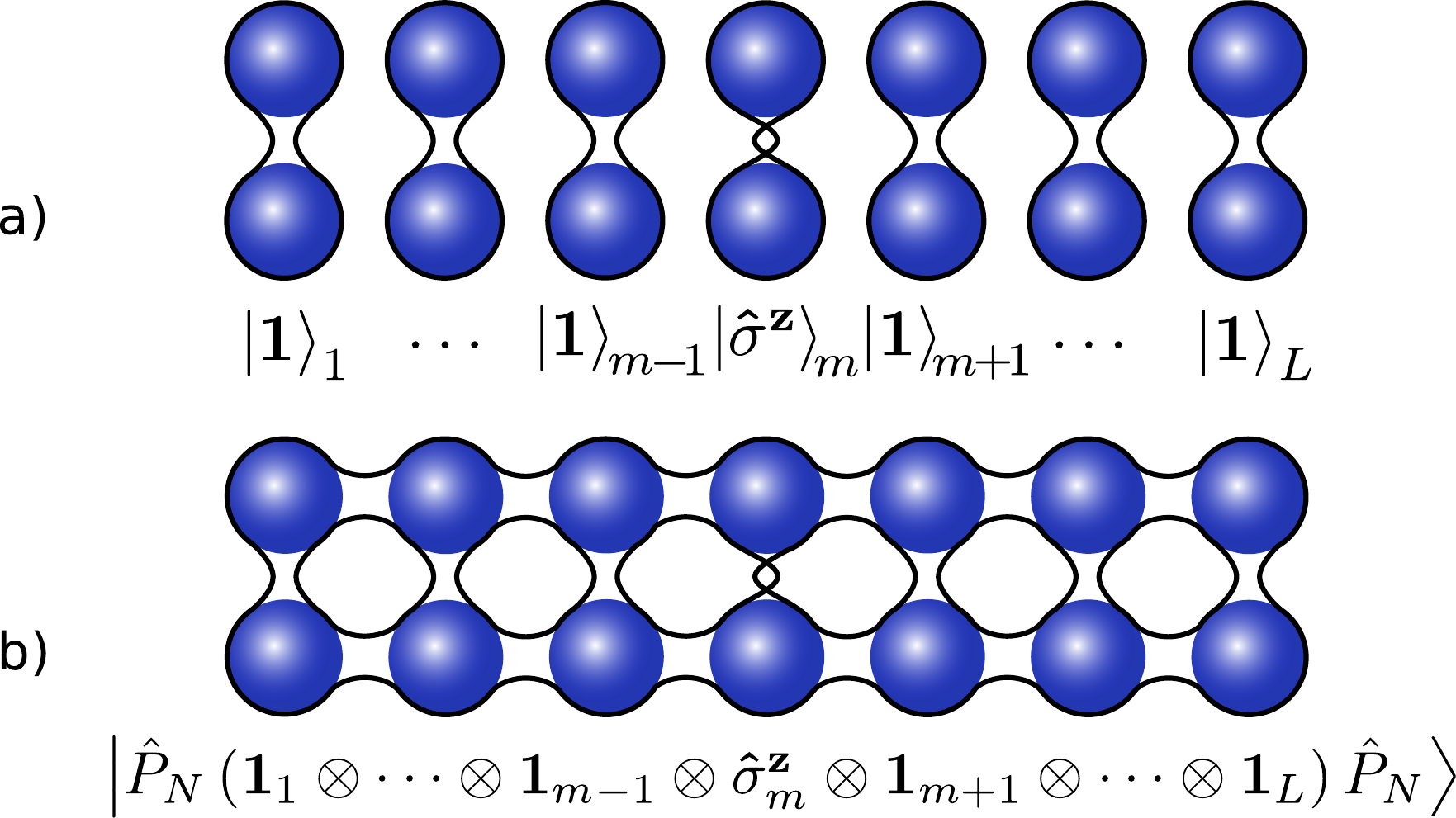}
\caption{a) In the grand canonical Hilbert space a product operator can be interpreted as a state on a double chain, which can be in general maximally entangled locally, but not at all along the chain. b) If the operator is however projected to a certain particle number, the corresponding double-chain state gets entangled also along the chain. This entanglement can overcompensate the benefits from shrinking the Hilbert space, depending on the actual particle number in question.}
\label{fig:entangled}
\end{figure*}

The purpose of the present work is to show how symmetries can be imposed on MPOs in general and to discuss the computational benefits and penalties. For simplicity, we will restrict the discussion to particle number conservation. It can be incorporated into MPOs on two levels: The first option reduces the Hilbert space dimension only halfway, as the operator is not projected onto a certain symmetry sector. It only requires the operator to annihilate (or create) a fixed number $\Delta N$ of particles (which might be zero),
\begin{equation}
 \hO = \sum_{N} \hat{P}_{N-\Delta N}\hO\hat{P}_N,
\label{eq:projsDelta}
\end{equation}
where $\hat{P}_N$ is the projectors onto the $N$ particle Hilbert space. This property is conserved under time-evolution with a particle-number conserving Hamiltonian. The operator is not restricted to any particular input particle number. We will therefore refer to this as the grand-canonical method\footnote{We do however not require the operator to actually be a density matrix, which would imply $\Delta N = 0$}.The way the symmetry is imposed is then equivalent to the usual way of adjoining good quantum numbers to an MPO\cite{Schollwock2011}. This approach has already proven useful in practical calculations\cite{Muth2011} and introduces no entanglement overhead. With the second option, we however go a step further: The Hilbert space dimension is reduced further by projecting onto a symmetry sector. This second method of using the conservation law for MPOs restricts the operator to a particular input particle number. We will therefore refer to it as the canonical method. This approach however introduces additional entanglement in the MPO, as illustrated in figure \ref{fig:entangled}. If the filling (number of particles per lattice site) is sufficiently low this is not a problem. However it can make the method less useful in the generic case.

The structure of this paper is as follows: Section \ref{sec:mps} gives a brief review of MPS and how a given symmetry can be explicitly accounted for. Section \ref{sec:mpo} introduces MPO and the interpretation of an MPO as an MPS in the ``super-space'' of operators. In sections \ref{sec:ur} and \ref{sec:r} we will discuss the two distinct ways of imposing particle umber conservation onto MPOs in detail. Section \ref{sec:fock} will give details on how to construct the projected operator in practise and gives exact results on the entanglement overhead introduced. Section \ref{sec:ex} gives example calculations, which illustrate how the two methods and also the brute force method, where no symmetry is taken into account, perform in comparison.

\section{Matrix product states}
\label{sec:mps}

MPS are an efficient way of specifying the state
\begin{eqnarray}
 \ket{\Psi} &=& \sum_{\vec{j}} c_{\vec{j}} \ket{\vec{j}}
\end{eqnarray}
(assumed here to be normalised) of a one dimensional lattice system. Here the $\vec{j}$ is a vectors of occupation numbers (or whatever other quantities are required to uniquely define the state of a single site) for every lattice site, thus corresponding to a Fock state. The number of parameters $c_{\vec{j}}$ is exponentially large in the system size. An MPS reduces this number by parametrising the state in terms of finite size matrices $A$, which we will assume here to be all square and of dimension $\chi\times\chi$:
\begin{eqnarray}
 \ket{\Psi} &=& \sum_{\vec{j}} c_{\vec{j}} \ket{\vec{j}} = \sum_{\vec{j}} \Tr{\left[\prod_m A^{[m],j_m}\right]} \ket{\vec{j}} \label{eq:mps1}\\
 &=& \Tr{\left[\bigotimes_m  \left(\sum_{j}A^{[m],j} \ket{j}_m\right)\right]} \label{eq:mps2}
\end{eqnarray}
While the product $\prod$ in (\ref{eq:mps1}) denotes the usual matrix product, $\bigotimes$ in (\ref{eq:mps2}) denotes matrix product between $A$ matrices and at the same time the direct product of the states of each lattice site. $\Tr$ here means taking the trace over the auxiliary space, i.e., the one where the matrices $A$ act, only. The trace is required only for systems with periodic boundary conditions, while for finite systems comprising $L$ sites the matrices belonging to the first site ($A^{[1],j}$) and those belonging to the last site ($A^{[L],j}$) all can be chosen to be row respectively column vectors instead of matrices. Equation (\ref{eq:mps2}) shows that an MPS is a generalisation of the notion of a usual product state, to which it reduces if all $A^{[m],j}$ are complex numbers, i.e., $\chi=1$.

Obviously the matrices $A$ are not uniquely defined by (\ref{eq:mps1}). If the system is not subject to periodic boundary conditions, there is however a unique canonical\cite{Perez-Garcia2007} form of the MPS (In this context the word ``canonical'' refers to the orthogonality and normalisation properties which we do not find in an MPS in its general form (\ref{eq:mps2}) and is not to be mixed up with canonical in the sense of working at a fixed particle number, section \ref{sec:r}.):
\begin{equation}
 \ket{\Psi} = \sum_{\vec{j}}
 \Gamma^{[1]j_1} \lambda^{[1]} \cdots \lambda^{[m-1]} \Gamma^{[m]j_m} \lambda^{[m]} \cdots \lambda^{[L-1]} \Gamma^{[L]j_L} \ket{\vec{j}}
\label{eq:mpscan}
\end{equation}
The $\lambda^{[m]}$ matrices are diagonal and contain the singular values from a Schmidt decomposition of a bi-partition of the system into the sub-chain A comprising sites $1$ to $m$ and the sub-chain B comprising sites $m+1$ to $L$ (in descending order for uniqueness):
\begin{equation}
 \ket{\Psi} = \sum_{\alpha=1}^\chi \lambda^{[m]}_\alpha \ket{\alpha}_{\textrm A} \otimes \ket{\alpha}_{\textrm B}
\label{eq:mpsSchmidt}
\end{equation}
$\left\{ \ket{\alpha}_{\textrm A} \right\}$ and $\left\{ \ket{\alpha}_{\textrm B} \right\}$ respectively form an orthonormal set, the reduced bases.

From this representation the constraint of MPS becomes apparent: The maximum number of nonzero singular values is $\chi$. For a general state, this number can be the smaller of Hilbert space dimensions of A and B. At the heart of DMRG lies the discarding of all but the $\chi$ largest eigenvalues of the reduced density matrix of any of the two subsystems, which is equivalent to approximating the state by an MPS with dimension $\chi$. Roughly speaking, this approximation is only good, if the entanglement entropy
\begin{equation}
 S^{[m]} = -\sum_{\alpha=1}^\chi \left.\lambda^{[m]}_\alpha\right.^2\log_2\left( \left.\lambda^{[m]}_\alpha\right.^2\right)
\label{eq:entropy}
\end{equation}
between A and B is small. (Rigorous results on the approximability in terms of entanglement entropies can be found in \cite{Schuch2008}.) Low lying eigenstates of 1D systems with short-range interaction can be approximated well \cite{Verstraete2006}, as $S$ grows only logarithmically with system size. In real-time evolution however, $S$ in general grows linear in time \cite{Calabrese2005}, restricting t-DMRG to short times.

If the state $\ket{\Psi}$ is an eigenstate of the total particle number in the whole system with eigenvalue $N$, then the Schmidt vectors $\ket{\alpha}_{\textrm A}$ and $\ket{\alpha}_{\textrm B}$ also have to be eigenstates of the total particle number in there respective subsystems, their eigenvalues $N_{\textrm A}(\alpha)$ and $N_{\textrm B}(\alpha)$ adding up to $N$ but maybe different for different values of $\alpha$.

The Schmidt decomposition at two neighbouring bonds reads
\begin{equation}
 \ket{\Psi} = \sum_{\alpha,\beta=1}^\chi \sum_{j}
 \ \lambda^{[m-1]}_\alpha \Gamma^{[m]j}_{\alpha,\beta} \lambda^{[m]}_\beta
 \ \ket{j}_m \otimes \ket{\alpha}_{\textrm A} \otimes \ket{\beta}_{\textrm B}.
\label{eq:mpssdplain}
\end{equation}
Here A comprising sites $1$ to $m-1$. If the state has a certain symmetry, this restricts the number of allowed states $\ket{j}_m$ for given $\ket{\alpha}_{\textrm A}$ and $\ket{\beta}_{\textrm B}$. In the case of particle number conservation we have $j=N-N_{\textrm A}(\alpha)-N_{\textrm B}(\alpha)$. When implementing this scheme, we can therefore leave out the physical dimension of the tensor $\Gamma^{[m]j}_{\alpha,\beta}$ completely (if there is no further local degree of freedom besides occupation number). This is one point where the conservation law makes the algorithm more efficient in terms of memory. The tensor $\Gamma^{[m]j}_{\alpha,\beta}$ is said to be symmetric. For a more mathematical description in terms of group theory see, e.g., \cite{McCulloch2007, Bauer2011} and references therein. It should be noted at this point, that applying particle number conservation in this way to MPS for infinite size systems \cite{Vidal2007} requires the average filling to be a multiple of one over the size of the unit cell\cite{McCulloch2008}, thus in general requiring large unit cells to approximate generic filling.

The only nontrivial (i.e., not conserving the MPS structure automatically) operation required to perform calculations, e.g. using the TEBD scheme\cite{Vidal2003, Vidal2004}, is acting with an operator on two neighbouring sites. After applying a particle number conserving operator ($U^{ij}_{i'j'}=0$ if $i+j\ne i'+j'$),
\begin{eqnarray}
 \hU \ket{\Psi} = \sum_{\alpha,\beta=1}^\chi \sum_{i, j} \sum_{i', j'} &\ & U^{ij}_{i'j'} \cdot
 \lambda^{[m-1]}_\alpha \Gamma^{[m]i'}_{\alpha,\beta} \lambda^{[m]}_\beta \Gamma^{[m]j'}_{\beta,\gamma} \lambda^{[m]}_\gamma \times \nonumber\\
 &\times&\ \ket{i}_m  \otimes \ket{j}_{m+1} \otimes \ket{\alpha}_{\textrm A} \otimes \ket{\gamma}_{\textrm B},
\label{eq:mpsneighbour}
\end{eqnarray}
here B comprising sites $m+2$ to $L$, the singular value decomposition of the tensor
\begin{equation}
 T^{i\alpha}_{j\gamma} = \sum_{i', j', \beta} U^{ij}_{i'j'} \Gamma^{[m]i'}_{\alpha,\beta} \lambda^{[m]}_\beta \Gamma^{[m]j'}_{\beta,\gamma}
\label{eq:T}
\end{equation}
has to be calculated, to bring the MPS back to it's canonical form. However it will be composed of blocks, each having a fixed value of $N_{\textrm A}(\alpha)+i$ (and correspondingly $j+N_{\textrm B}(\alpha)=N-N_{\textrm A}(\alpha)-i$). Typically these blocks are much smaller each then the size of $T$ itself, and can moreover be decomposed in parallel, such that the fixed symmetry gives a big advantage when operating on MPS.

We note that the case of non abelian symmetries \cite{McCulloch2002, Singh2010, Schollwock2011} is more involved and will not be discussed here.

\section{Matrix product operators}
\label{sec:mpo}

MPOs can be understood as a generalisation of product operators in the same way as shown in (\ref{eq:mps2}) for MPS,
\begin{eqnarray}
 \hO &=& \Tr{\left[\bigotimes_m \left(\sum_{j} A^{[m],j} \ho_j\right)\right]}. \label{eq:mpo}
\end{eqnarray}
The $\ho_j$ form a orthogonal basis of the local operator space of single site, normalised according to the Hilbert-Schmidt norm,
\begin{equation}
 \Tr\left[\hod_j\ho_j'\right] = \delta_{j, j'},
\end{equation}
where the trace is here over the physical space.

The space of operators can be mapped to a ``super-space'' of kets via
\begin{equation}
 \hO = \sum_{\vec{i},\vec{j}} o_{\vec{i},\vec{j}} \ket{\vec{i}}\bra{\vec{j}}\ \longmapsto\ \ket{\hO} = \sum_{\vec{i},\vec{j}} o_{\vec{i},\vec{j}} \Bigket{\begin{array}{c}\vec{j}\\\vec{i}\end{array}}.
\label{eq:superstate}
\end{equation}
Again the $\vec{i}$ and $\vec{j}$ are vectors of occupation numbers for every lattice site, thus corresponding to a Fock state. In these terms we talk about an upper in- and a lower out-chain. (``In'' and ``out'' refer to the original operator acting as a function.) An MPO is then equivalent to an MPS representation of such a ``super-state''. E.g., the Schmidt decomposition at two neighbouring bonds reads
\begin{equation}
 \ket{\hO} = \sum_{\alpha,\beta=1}^\chi \sum_{i=0}^{d-1}  \sum_{j=0}^{d-1}
 \lambda^{[m-1]}_\alpha \Gamma^{[m]i,j}_{\alpha,\beta} \lambda^{[m]}_\beta
 \Bigket{\begin{array}{c}j\\i\end{array}}_{m} \otimes \ket{\alpha}_{\textrm A} \otimes \ket{\beta}_{\textrm B}.
\label{eq:sdplain}
\end{equation}
The structure of the $\Gamma$ tensors for certain symmetric operators will be discussed in sections \ref{sec:ur} and \ref{sec:r}. Note that the local Hilbert space dimension $d$ in general has to be restricted to some reasonable value, usually by allowing for a maximum on-site particle number $d-1$. This is because otherwise certain operators (even such basic ones as a particle annihilator or even unity) would have non-vanishing contributions from infinitely many particle numbers and in general can not even be normalised.

The relevant measure for the resources required to approximate an operator well by an MPO is now the entanglement in operator space, a possible measure being the operator space entanglement entropy \cite{Prosen2007} (OSEE) which is defines just as the entanglement entropy (\ref{eq:entropy}) for MPS. It must not be confused with the systems statistical entropy when interpreting $\hO$ as a density matrix: As a striking example, the infinite temperature density matrix $\mathbf{1}/d^{L}$ has maximal statistical entropy of $L\log_2(d)$ but it is clearly a product operator and therefor the OSEE is $0$. On the other hand a projector $\ket{\Psi}\bra{\Psi}$ is always pure and has statistical entropy $0$ while its OSEE is just the entanglement entropy of the state $\ket{\Psi}$ which can be as large as $L\log_2(d)/2$, in which case there will be no efficient approximation by an MPO.

We observe that a product operator always maps to a product state, i.e., with bond dimension $\chi=1$, and therefore with no entanglement between the sites:
\begin{equation}
\label{eq:prodop}
 \bigotimes_m \hO^{[m]}
 = \bigotimes_m \left( \sum_{i,j}  o^{[m]}_{j,i} \ket{i}_m\bra{j}_m \right)
\ \longmapsto\ 
 \bigotimes_m \left( \sum_{i,j} o^{[m]}_{j,i} \Bigket{\begin{array}{c}j\\i\end{array}}_m\right).
\end{equation}
Product operators, and eventually sums of a small number of those, e.g. correlators, form the majority of physically interesting quantities, namely those which are potentially measurable in real many body systems. What makes a Heisenberg picture simulation a promising approach, is the fact that we find no entanglement in them in the first place, namely at time $t=0$. If however we project a product operator to a given symmetry sector, as discussed in detail in section \ref{sec:r}, it does not necessarily map onto a product state any more. This is illustrated in figure \ref{fig:entangled}.

One generally expects the OSEE to grow linear with time in a dynamical simulation in the Heisenberg picture. However it has been conjectured\cite{Prosen2007a, Prosen2007} that this scaling becomes logarithmical and (therefore allowing for efficient classical simulation using a t-DMRG scheme) for integrable models. Recently it has been verified, that already a single conserved quantity (like the total number of particles) is sufficient to guarantee such favourable scaling for certain operators \cite{Muth2011}. We will therefore put the focus on operators that are typical observables, propagated in Heisenberg picture simulations.

The Heisenberg equation of motion for the operators, $i \partial_t \hO = \ko{\hO}{\hH}$, gives rise to a Schr\"odinger type equation of motion for the ``super-states``, $i\partial_t\ket{\hO} = \tH\ket{\hO}$, with the new Hamiltonian
\begin{equation}
 \tH = \mathbf{1} \otimes \hH - \hH \otimes \mathbf{1}.
\end{equation}
Thus this ``super-Hamiltonian'' acts on the in- and out-chains independently. (In general the dynamics is however not just the dynamics of two independent chains, because the initial operator will be mapped to a state with strong entanglement between in- and out-chain, equation (\ref{eq:superstate}).) If $\hH$ conserves the total particle number  $\hN = \sum_m \hn_m$, i.e., $\ko{\hN}{\hH}=0$, there exist two different ways of constructing MPOs that take advantage of this.

\section{Unprojected operators}
\label{sec:ur}

To introduce the first method, which works in the grand-canonical Hilbert space, we observe that, because $\hH$ conserves the total particle number, $\tH$ conserves the number \emph{difference} between the in- and out- chains:
\begin{equation}
 \ko{\tH}{\hN \otimes \mathbf{1} - \mathbf{1} \otimes \hN} = 0
\end{equation}

If we consider operators (\ref{eq:projsDelta}) which annihilate (or create) a fixed number of particles $\Delta N$
(like, e.g., the particle annihilation and creation operators $\ha_m$ or $\had_m$ at some given site $m$, and products of those) they will map to an eigenstates of this difference. (If $\hO$ is a density matrix, $[\hO, \hN]=0$ and therefore it is such an operator and $\Delta N=0$.) Note that the identity on the whole Hilbert space,
\begin{equation}
 \mathbf{1} = \bigotimes_m \mathbf{1}^{[m]} = \sum_{N} \hat{P}_N,
\end{equation}
is a prototype of such an operator. $\Delta N$ now is a conserved quantity in the super-space. The Heisenberg dynamics will then take place only in a specific symmetry sector (with a fixed $\Delta N$) and the MPO can be restricted accordingly. This can be done in exactly the same way as for MPS when the total number itself is conserved:

Given the canonical form of the MPO and the Schmidt decomposition at two neighbouring bonds, equation (\ref{eq:sdplain}). If $\ket{\hO}$ is a particle number difference eigenstate, then also the Schmidt vectors $\ket{\alpha}_{\textrm A}$ and $\ket{\beta}_{\textrm B}$ must be eigenstates of the particle number difference in their respective subsystems. The local particle number difference $j-i$ is thus determined from $j$, $\alpha$ and $\beta$ alone,
\begin{eqnarray}
 j+\Delta N_{\textrm B} = i-\Delta N_{\textrm A} +\Delta N.
\end{eqnarray}
In an actual implementation (with no further local degrees of freedom) there is no second physical index $i$ necessary in the $\Gamma$ tensor. This particular form of the MPS can be kept during time evolution, using the scheme discussed in section \ref{sec:mps}. For details we refer the reader to the literature \cite{Daley2005a, McCulloch2007, Singh2010}.

This grand-canonical method gives great advantage over the plain approach\cite{Muth2011} which works for general systems without conservation laws. A comparison for an example case can be found in section \ref{sec:ex}.

Before we continue to the second method, we take a look at the entanglement in this first approach. Therefore we give an explicit construction of the initial MPO in two steps. The first step is the construction of an MPO for the identity. This task is trivial, but we take a route that can be conveniently generalised in section \ref{sec:r}: Given a state $\sum_{\vec{j}} c_{\vec{j}} \ket{\vec{j}}$, that is a superposition of Fock states, the mapping
\begin{equation}
 \ket{j}_m \ \longmapsto\  \ket{j}_m\bra{j}_m,
 \label{eq:mpsmpo}
\end{equation}
which is applied locally at every site $m$ simultaneously, maps it to a superposition $\sum_{\vec{j}} c_{\vec{j}} \ket{\vec{j}}\bra{\vec{j}}$ of projectors onto these Fock states. We get the identity matrix by superimposing \emph{all} Fock states with amplitude $c_{\vec{j}}=1$,
\begin{equation}
 \sum_{\vec{j}} \ket{\vec{j}} = \bigotimes_m \left(\sum_{j=0}^{d-1} \ket{j}_m \right)
\ \longmapsto\ 
 \bigotimes_m \left(\sum_{j=0}^{d-1} \ket{j}_m\bra{j}_m \right) = \sum_{\vec{j}} \ket{\vec{j}}\bra{\vec{j}} = \mathbf{1}.
\end{equation}
It's MPO representation,
\begin{eqnarray}
 \ket{\mathbf{1}} &=& \sum_{\vec{j}} \ket{\begin{array}{c}\vec{j}\\\vec{j}\end{array}} = \bigotimes_m \left(\sum_{j=0}^{d-1} \Bigket{\begin{array}{c}j\\j\end{array}}_m \right) 
\label{eq:erklaer}
\end{eqnarray}
thus has large entanglement between the two chains. The entanglement is however contained \emph{within} the matrices themselves. There is no entanglement between different lattice sites, thus a bond dimension of $\chi=1$ suffices. The matrices of the MPO are simply $1$.

In the second step we get the MPO representation of $\hO$ by \emph{applying} $\hO$ itself only to the out-chain of $\ket{\mathbf{1}}$,
\begin{equation}
 \ket{\hO} = \sum_{\vec{i},\vec{j}} o_{\vec{i},\vec{j}} \Bigket{\begin{array}{c}\vec{j}\\\vec{i}\end{array}} = \left(\mathbf{1}\otimes\hO\right)\ket{\mathbf{1}}.
\label{eq:1toOgc}
\end{equation}
A typical observable will be reasonably simple, e.g., a two point correlator $\had_m\ha_m\had_{m'}\ha_{m'}$ which is a product operator or a local current $i\left(\had_j\ha_{j+1}-\had_{j+1}\ha_j\right)$ for which $\chi=2$. Then its MPO will also have a simple form. This will change dramatically however, if we project the operator to the subspace of a given particle number, as discussed in the next section.

\section{Projected operators}
\label{sec:r}

This method works in the canonical Hilbert space. $\tH$ does of course not only conserve the number difference between the two chains, but also the total numbers in in the in-chain, $\hN^{({\textrm{in}})}=\hN\otimes\mathbf{1}$, and in the out-chain, $\hN^{({\textrm{out}})}=\mathbf{1}\otimes\hN$, \emph{separately}:
\begin{equation}
 \ko{\tH}{\hN \otimes \mathbf{1}}= \ko{\tH}{\mathbf{1} \otimes \hN} = 0
\end{equation}
However MPO representations for general operators (\ref{eq:projsDelta}) are not eigenstates of any of these. Taking into account particle number conservation in each chain separately therefore only applies to operators that are nonzero only in a given symmetry sector. If we project the operator to a given input particle number $N$, i.e., take only one of the summands in (\ref{eq:projsDelta}), 
\begin{equation}
  \hO_N =  \hat{P}_{N-\Delta N}\hO\hat{P}_N,
\label{eq:projsDeltaSingle}
\end{equation}
we find an eigenstate of the total particle number in the upper \emph{and} in the lower chain simultaneously. Thus when working in the canonical Hilbert space, particle number conservation can be used twice:

If in (\ref{eq:sdplain}) $\ket{\hO}$ is a particle number eigenstate in both chains, then also the Schmidt vectors $\ket{\alpha}_{\textrm A}$ and $\ket{\beta}_{\textrm B}$ must be eigenstates of the particle number in both chains in their respective subsystems. The local particle numbers $i$ and $j$ are thus determined from $\alpha$ and $\beta$ alone,
\begin{eqnarray}
 N^{({\textrm{in}})}_{\textrm A} + j + N^{({\textrm{in}})}_{\textrm B} &=& N \nonumber\\
 N^{({\textrm{out}})}_{\textrm A} + i + N^{({\textrm{out}})}_{\textrm B} &=& N-\Delta N.
\end{eqnarray}
In an actual implementation (with no further local degrees of freedom) there are no physical indices $i$ and $j$ at all necessary in the $\Gamma$ tensor. This particular form of the MPS can again be kept during time evolution. Thereby the $T$ tensor, equation (\ref{eq:T}), will break up into even smaller blocks, speeding up the calculation of it's singular value decomposition even more than in the grand-canonical method.

We get the MPO representation of $\hO_N$ by applying $\hO$ itself to the out-chain of the MPO representation of $\hat{P}_N$,
\begin{equation}
 \ket{\hO_N} = \ket{\hat{P}_{N-\Delta N}\hO\hat{P}_N} = \left(\mathbf{1}\otimes\hO\right)\ket{\hat{P}_N}.
\end{equation}
$\hat{P}_N$ is now the identity \emph{only in the sector of particle number $N$}. It vanishes in the remains of the grand canonical Hilbert space. $\hat{P}_N$ takes the role as a prototype of a projected operator, analogous to identity in (\ref{eq:1toOgc}). The difficulty of the second approach results from the fact that $\hat{P}_N$ is clearly not a product operator, but entangled between the sites, as illustrated in figure \ref{fig:entangled}. We will construct it explicitly in section \ref{sec:fock}.

Working with the canonical method has the advantage, that we do not have to limit the local dimension explicitly to $d$, as $d < N$ is automatically fulfilled, which comes in handy, e.g., for bosonic models.

Of course $\hO$ and $\hO_N$ are not equivalent. But in certain cases this is not relevant, e.g. if $\hO$ is an observable (which implies $\Delta N=0$), and we evolve $\hO$ in time using Heisenberg t-DMRG in order to find the dynamics of its expectation value. Then the result is the same using $\hO_N$ if the state $\ket{\Psi_0}$ of the system for which we want to calculate the expectation value is a particle number eigenstate, $\hN\ket{\Psi_0}=N\ket{\Psi_0}$:
\begin{equation}
 \bra{\Psi_0}\hO_t\ket{\Psi_0}
 = \bra{\Psi_0}\hat{P}_{N}\hO_t\hat{P}_N\ket{\Psi_0}
 = \bra{\Psi_0}\left(\hat{P}_{N}\hO\hat{P}_N\right)_t\ket{\Psi_0}
\end{equation}
An example of this type is given for a bosonic model at the end of section \ref{sec:ex}.

\section{Preparing the projector onto the subspace of a fixed particle number}
\label{sec:fock}

What is left is the construction of the MPO representation of $\ket{\hat{P}_N}$. Following the arguments in section \ref{sec:ur} for the construction of $\ket{\mathbf{1}}$, this reduces to preparing an MPS that is a superposition of all Fock states \emph{with total particle number} $N$, which will be discussed in the following. The operationally simple mapping (\ref{eq:mpsmpo}) together with (\ref{eq:superstate}) will transform it to $\ket{\hat{P}_N}$.

\begin{figure*}[htb!]
\centering
\includegraphics[width=.49\textwidth]{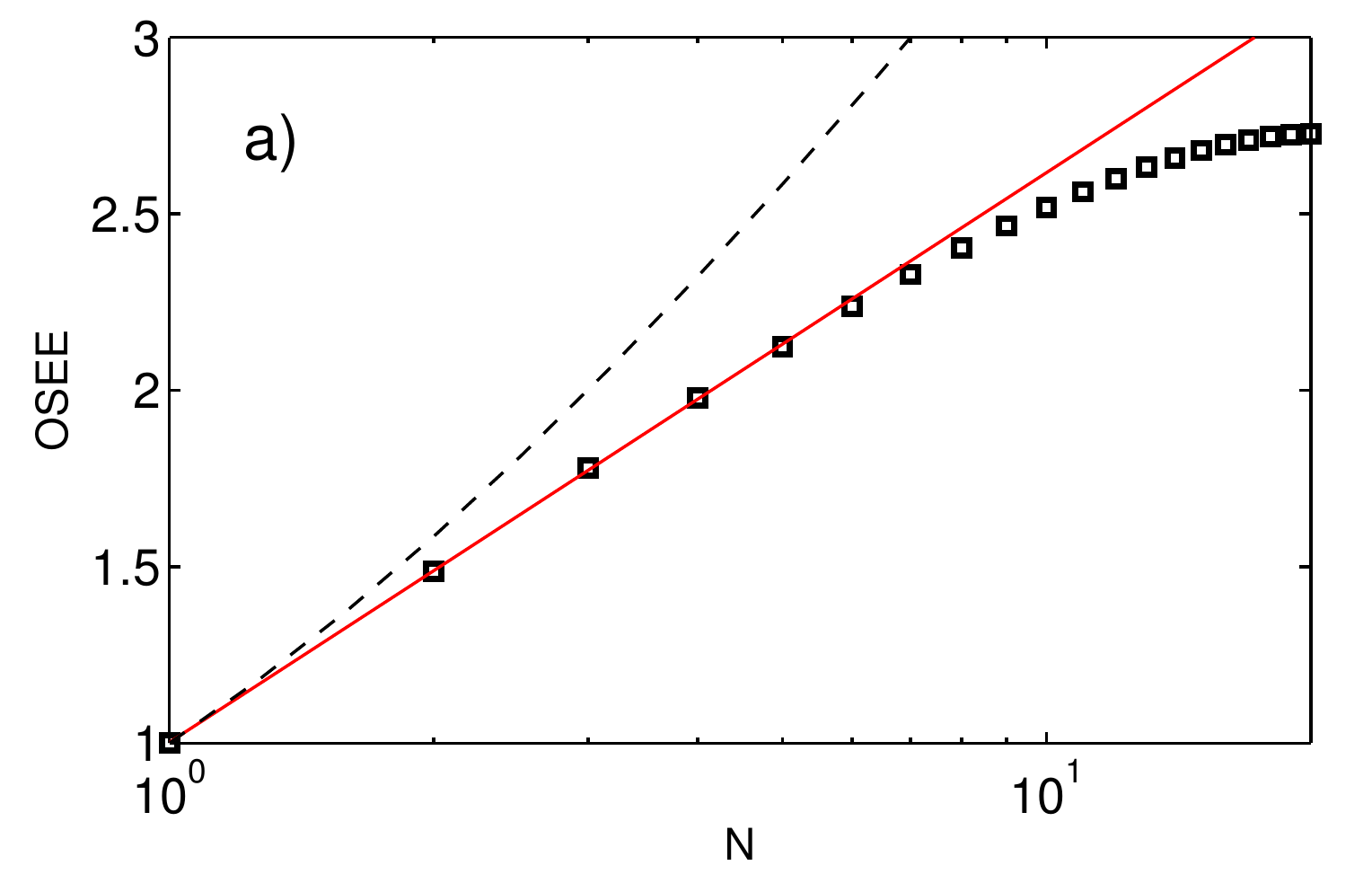}
\includegraphics[width=.49\textwidth]{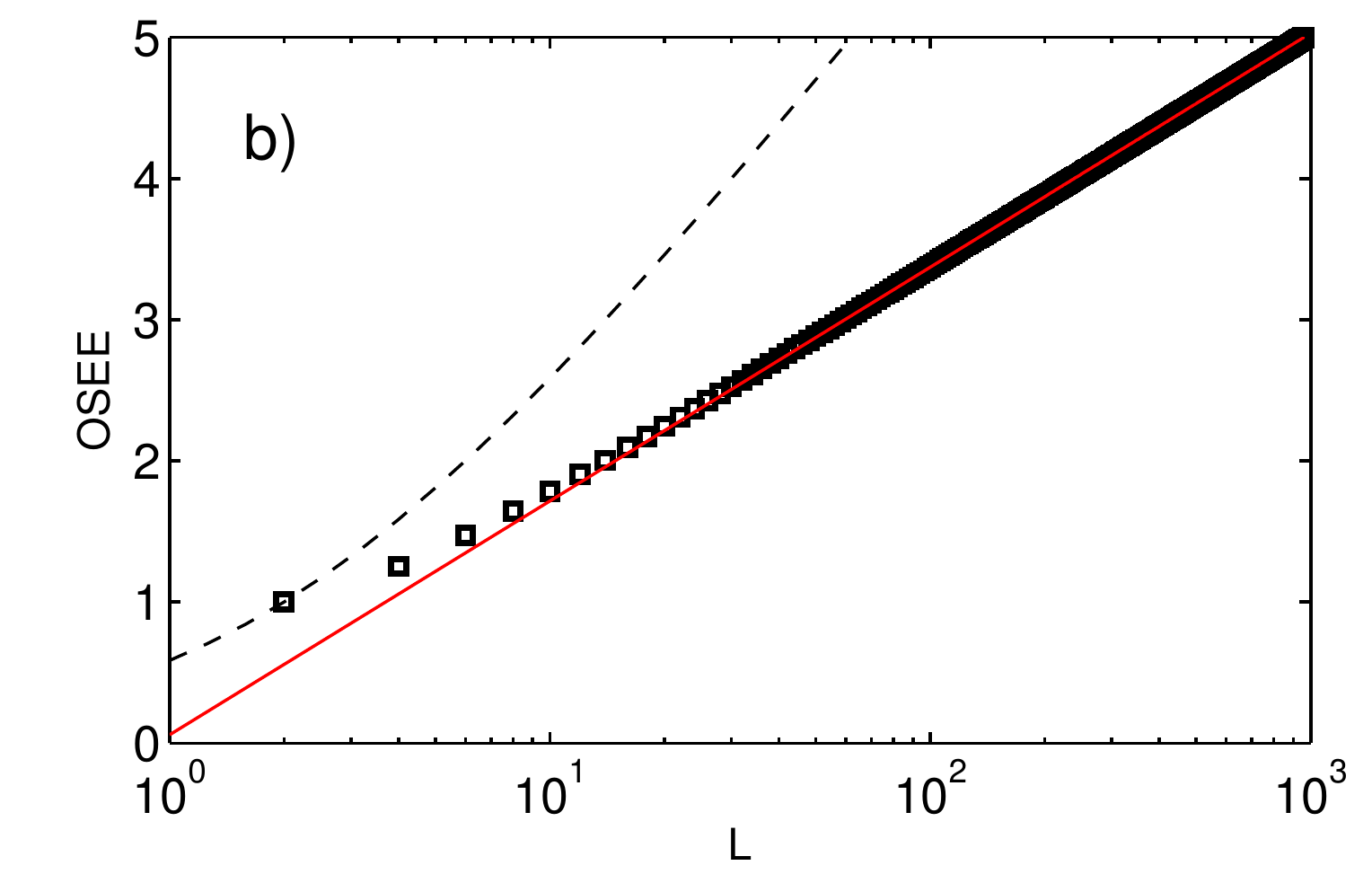}
\caption{Logarithmic scaling of the OSEE of the projector $\hat{P}_N$ at the centre of the chain with particle number $N$ and system size $L$. We show $S^{[L/2]}$ in a system of fermions ($d=2$) on a lattice, a) as a function of $N$ for fixed system size $L=40$. (Note that as $N > L/2$ the entropy goes down again due to the Pauli principle and particle hole symmetry.) b) as a function of $L$ for a fixed filling of $N/L = 1/2$. -- Symbols are from the numerical evaluation of (\ref{eq:lambda}). Straight lines show fits to these. Dashed lines show the upper limit (\ref{eq:oseelimit}). We checked numerically, that also for a higher local dimension $d$ the prefactor of the logarithmic scaling actually stays much below this limit.}
\label{fig:osee}
\end{figure*}

Let us denote the normalised, equal superposition of all $N$-particle Fock states which are locally constrained to a maximum particle number of $d-1$ by $\ket{N}$. If we want to work without a local constraint, we set $d=N+1$. Given a bi-partition of our lattice we note that its Schmidt decomposition is
\begin{equation}
 \ket{N} = \sum_{l=0}^N \lambda^{[m]}_l \ket{l}_{\textrm A} \otimes \ket{N-l}_{\textrm B}.
\end{equation}
The sub-chain A comprises sites $1$ to $m$, the sub-chain B comprises sites $m+1$ to $L$. This shows that the MPS will have bond dimension $\chi=N+1$. $\left.\lambda^{[m]}_l\right.^2$ is the probability of finding $l$ particles left of bond $m$:
\begin{equation}
 \left.\lambda_l^{[m]}\right.^2 = \frac{\Omega_{d}\left(l, m\right)\Omega_{d}\left(N-l, L-m\right)}{\Omega_{d}\left(N, L\right)}.
\label{eq:lambda}
\end{equation}
Here $\Omega_d(n, L)$ is the number of possibilities to distribute $n$ indistinguishable particles among $L$ sites in such a way that no site is occupied by more than $d-1$ particles, given by the recursion formula \cite{Freund1956}
\begin{equation}
 \Omega_d(n, L) = \sum_{j=0}^{\min(n, d-1)}\Omega_d(n-j, L-1);\quad \Omega_d(n, 0) = \delta_{n0}.
\end{equation}
For $d=2$ this reduces to $\Omega_2(n, L) = \binom{L}{n}$, for $d>n$ it reduces to $\Omega_{d>n}(n, L) = \binom{L+n-1}{n}$.

We continue with the Schmidt decomposition at the following bond. (Repeating it for all bonds leads to the canonical form of the MPS.) Here the remaining task is to determine the coefficients of
\begin{equation}
 \ket{N} = \sum_{l=0}^N \sum_{r=0}^N
 \lambda^{[m]}_l \Gamma^{[m+1]}_{lr} \lambda^{[m+1]}_r
 \ket{l}_{\textrm A} \otimes  \ket{r-l}_{m+1} \otimes \ket{N-r}_{\textrm B'}.
\end{equation}
The $\lambda$ tensors are already known from (\ref{eq:lambda}). The sub-chain ${\rm B}'$ comprises sites $m+2$ to $L$. Thus $\left.\Gamma^{[m+1]}_{lr}\right.^2\left.\lambda^{[m+1]}_r\right.^2$ is the probability of finding $N-r$ particles at the right side of bond $m+1$ \emph{provided that there are already $l$ particles at the left of bond $m$}:
\begin{eqnarray}
 \left.\Gamma^{[m+1]}_{lr}\right.^2\left.\lambda^{[m+1]}_r\right.^2
	&=& \frac{\Omega_d(r-l,1)\Omega_d(N-(r-l), L-1)}{\Omega_d(N, L)} \times \nonumber \\
	&\times& \frac{\Omega_d(l,m)\Omega_d(N-r, L-m-1)}{\Omega_d(N-(r-l), L-1)} \times \frac {1}{\left.\lambda^{[m]}_l\right.^2}.
\label{eq:gamma}
\end{eqnarray}

Equations (\ref{eq:lambda}) and (\ref{eq:gamma}) determine the $\Gamma$ and $\lambda$ tensors completely. Thus we can calculate the coefficients of the MPS exactly. By means of (\ref{eq:mpsmpo}) and (\ref{eq:superstate}) this also yields the MPO representation of $\hat{P}_N$:
\begin{equation}
 \ket{N}\ 
\begin{array}{c}(\ref{eq:mpsmpo})\\\longmapsto\\\phantom{1}\end{array}\ 
\frac{\hat{P}_N}{\sqrt{\Omega_d(N,L)}}\ 
\begin{array}{c}(\ref{eq:superstate})\\\longmapsto\\\phantom{1}\end{array}\ 
\frac{\ket{\hat{P}_N}}{\sqrt{\Omega_d(N,L)}}.
\end{equation}
 The $\Gamma$ matrices do not have physical indices explicitly, because the local particle numbers are given by the bond indices (due to particle number conservation), $l$ and $r$ here. Note that in this particular case, the value of the bond index has a physical meaning\footnote{I.e., here we have a one to one correspondence between index and good quantum number.}, namely the particle number at the left side of the bond $m$ considered. The absence of physical indices is especially useful for bosonic systems.

This construction shows the main difficulty of going to the canonical version of the MPO: Even the trivial operator $\mathbf{1}$ has an extensive bond dimension of $\chi = N+1$ if projected to a fixed particle number $N$. The initial entanglement, even of a local operator, is no longer only between the chains but also along the chain, see the illustration in figure \ref{fig:entangled} and the example in figure \ref{fig:cone}. Although computations can be done with a higher bond dimension here, this advantage is often overcompensated by the initial entanglement. However a linear growth of the matrix dimension does not imply, that the algorithm is inefficient. In contrast, the required matrix dimension in general grows exponentially with \emph{time}\cite{Calabrese2005}, which is a more severe limitation. Here, a linear scaling of the matrix dimension implies that the OSEE scales only logarithmically with system size,
\begin{equation}
 S^{[m]} = -\sum_{l=0}^L \left.\lambda^{[m]}_l\right.^2\log_2\left( \left.\lambda^{[m]}_l\right.^2\right)
 \le \log_2(N+1),
\label{eq:oseelimit}
\end{equation}
which is favourable\footnote{Actually the polynomial scaling of the matrix dimension can be taken as the definition of efficient. The nontrivial problem in general is to show that from the logarithmic scaling of the entanglement entropy one can conclude that there exist efficient approximations by MPS, see, e.g. \cite{Verstraete2006}.}. This entropy is minor compared to the entropy which has to be added on top for time evolution. Beyond that, for a low over all particle number $N$, the entropy that can be generated dynamically is bounded or at least drastically reduced. Longer times can then be reached as we will see in the example of the next section.

In fact the upper bound (\ref{eq:oseelimit}) is not even tight, as shown in figure \ref{fig:osee}. For details on the relation between the scaling of the entropy and the efficiency of an MPS see \cite{Schuch2008a}.

\section{Examples}
\label{sec:ex}

\begin{figure*}[htb!]
\centering
\includegraphics[width=.6\textwidth]{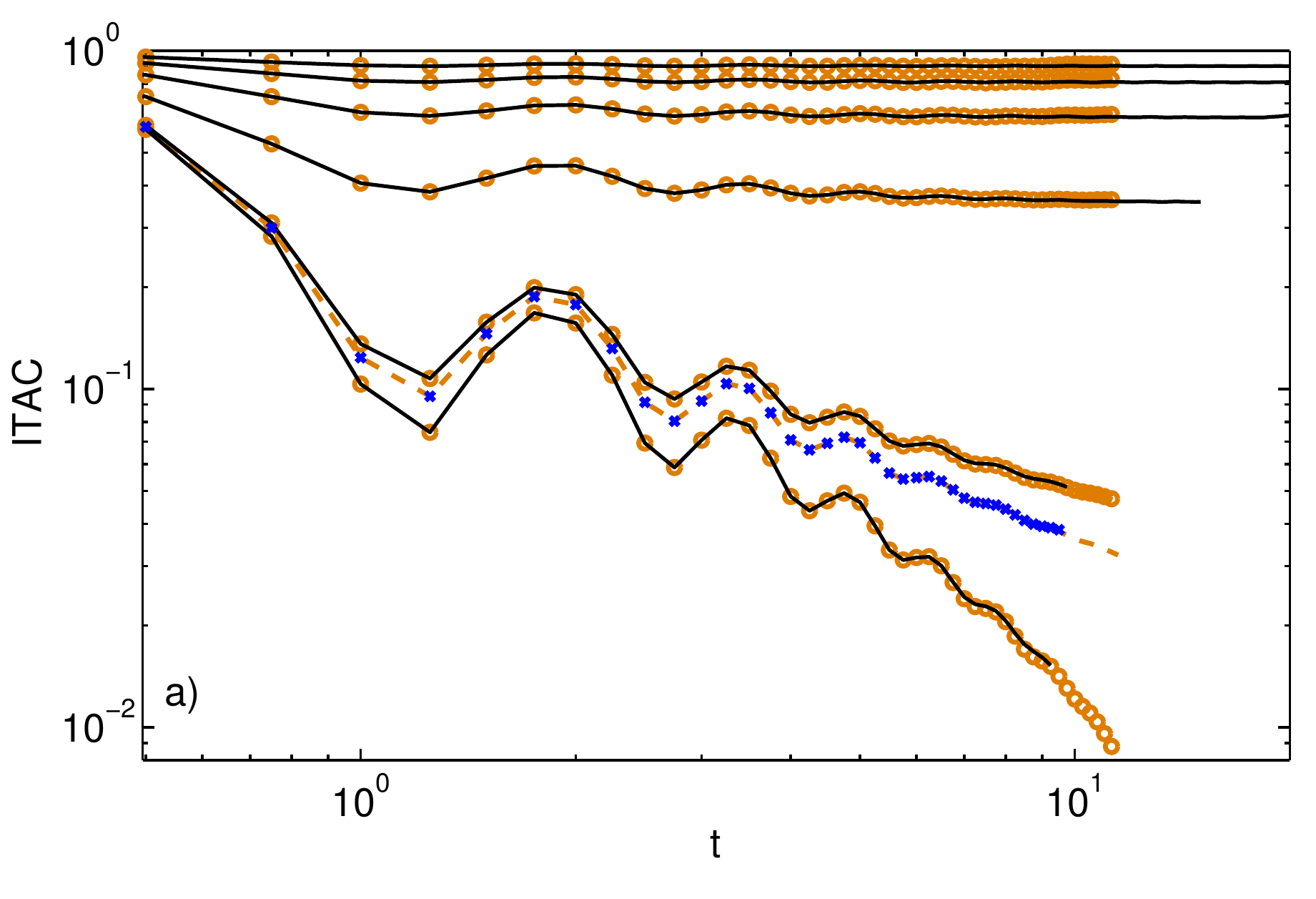}
\includegraphics[width=.6\textwidth]{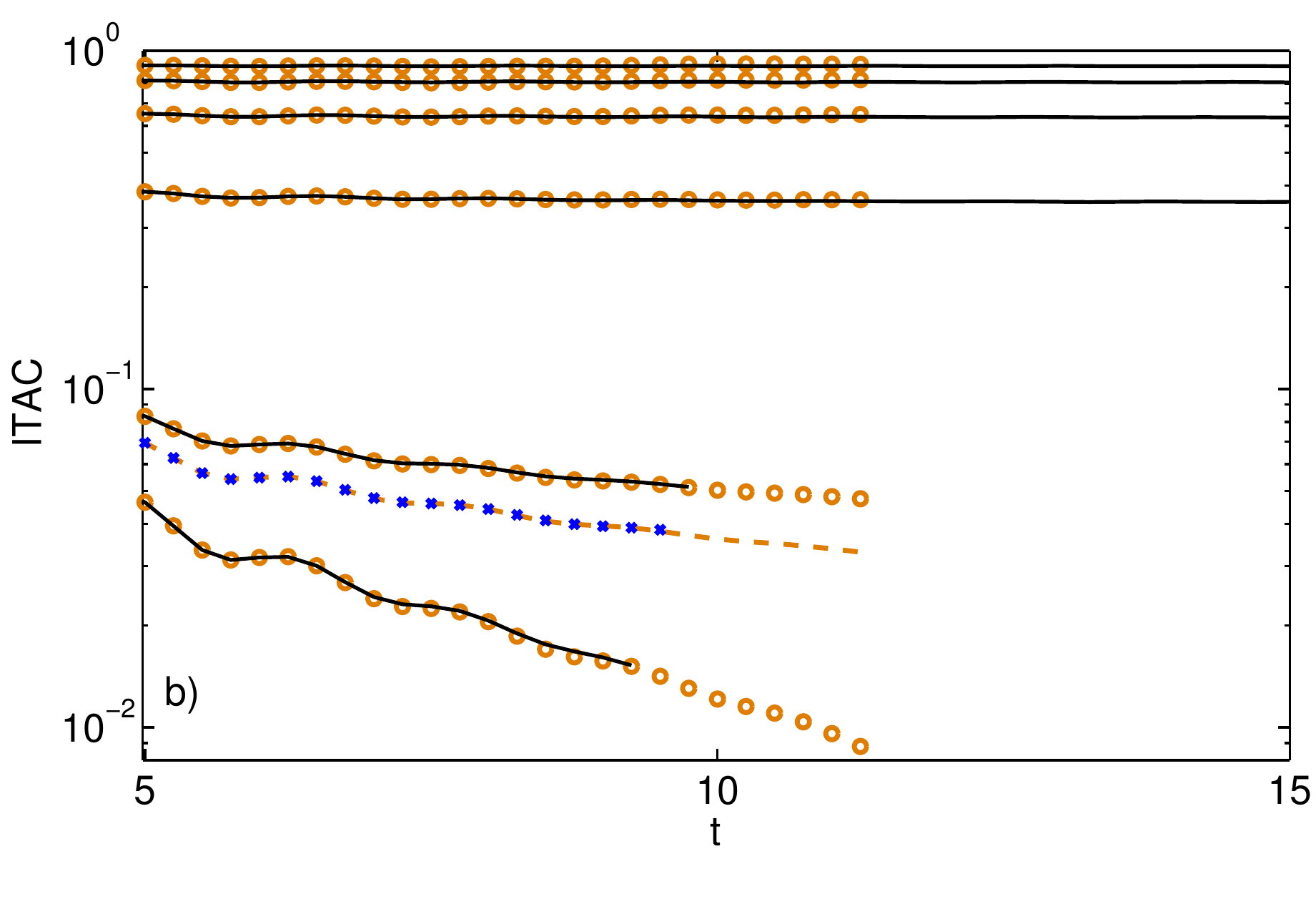}
\caption{
 Spin-$\frac12$ XXZ-chain of length $L=40$ at $\Delta=0.8$, time evolution in the Heisenberg picture. ITAC at site $m=20$: $\mathfrak{Re}\left[C_N(t)\right]$ ($N=1,2,4,8,16,20$ from top down, calculated using the canonical (solid line) and the grand-canonical method (orange circles)) and $\mathfrak{Re}\left[G(t)\right]$ calculated from the unprojected (dashed orange) and a brute force method (blue crosses) the latter ignoring particle number conservation completely. a) double logarithmic plot to emphasise the power law behaviour of the ITAC. b) same data as a), but linear time axis for better visibility of the difference in time reached by the different methods (and in the different symmetry sectors in case of the projected method). Bond dimensions used where $\chi=4000$ in the canonical calculations, $\chi=1000$ for the grand-canonical example, and $\chi=500$ in the brute force calculation. All curves end at the point where the accumulated cut-off error (see footnote on page \pageref{fn:cutoff}) reaches $10^{-2}$. A TEBD \cite{Vidal2003} version of the t-DMRG algorithm is used with a fourth order trotter decomposition and time step size $1/4$ in all cases. Curves are shown only up to $t=20$. At later times boundary effects show up, because the excitations have propagated to the end of the chain.
}
\label{fig:delta08}
\end{figure*}

\begin{figure*}[htb!]
\hspace*{2.3cm}\includegraphics[width=.6\textwidth]{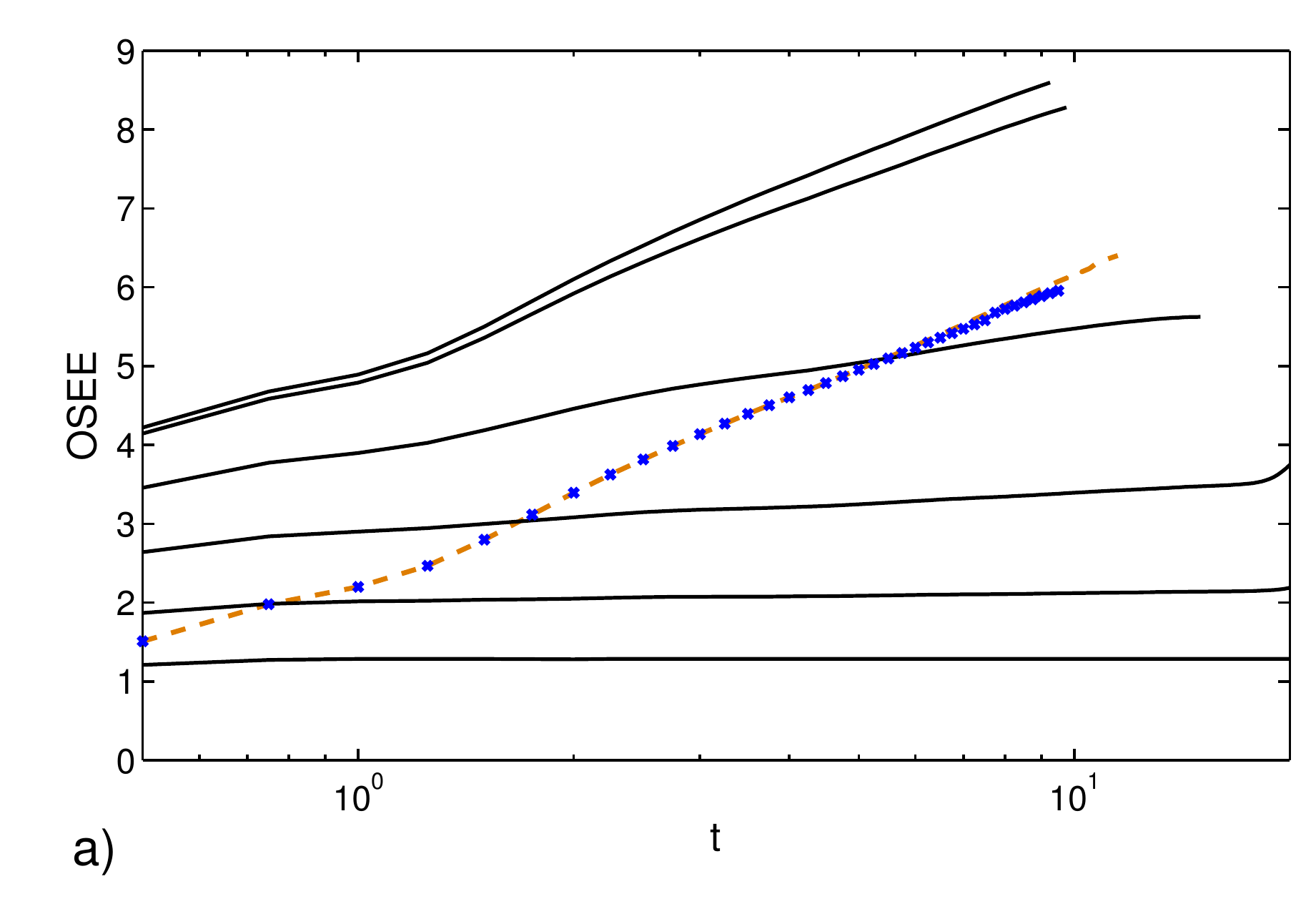}\\
\hspace*{2.63cm}\includegraphics[width=.7\textwidth]{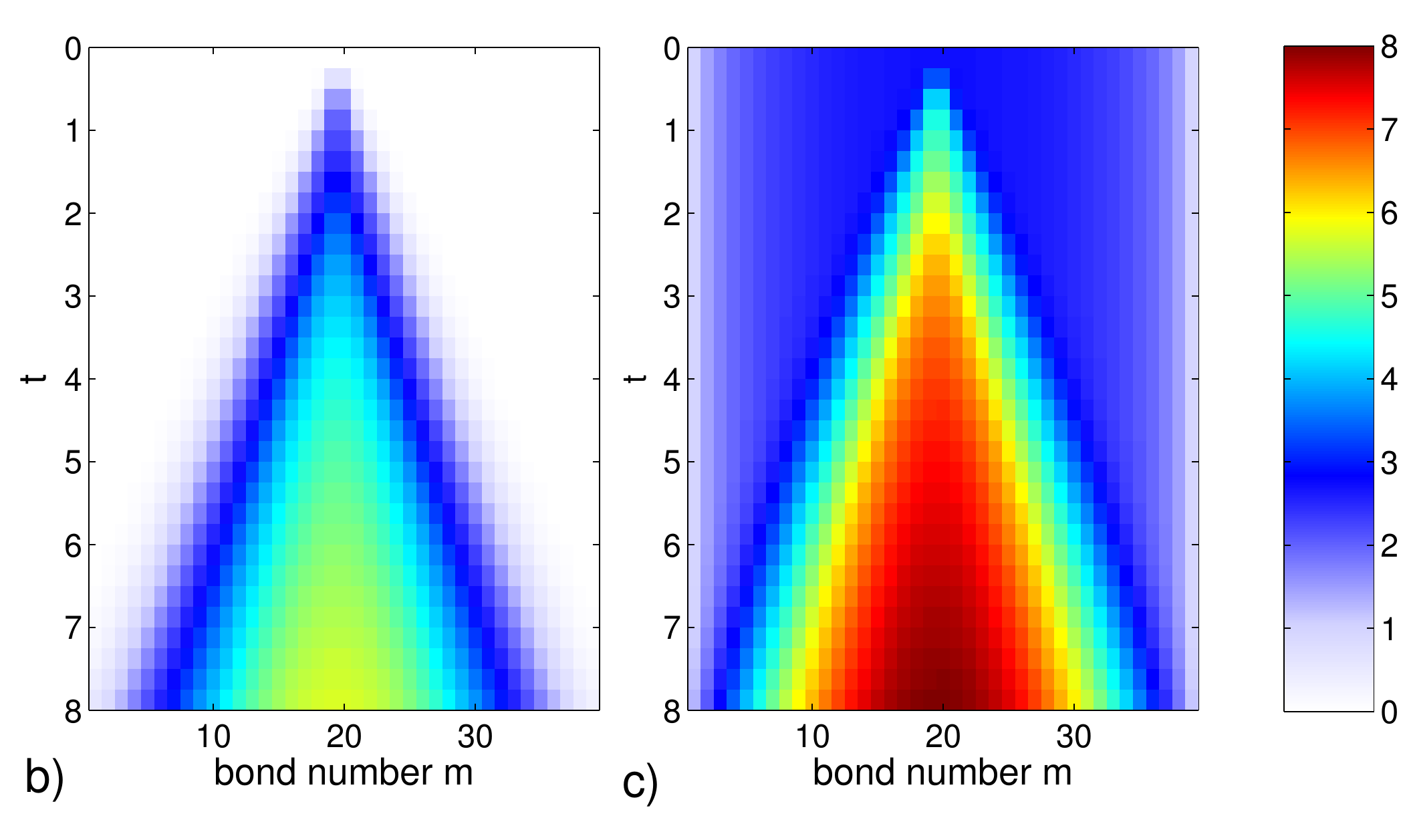}
\caption{
 Spin-$\frac12$ XXZ-chain of length $L=40$ at $\Delta=0.8$, time evolution in the Heisenberg picture. a) OSEE $S^{[20]}(t)$ of $\hat{P}_N\hsig^z_{20}\hat{P}_N$ (solid, $N=1,2,4,8,16,20$ from bottom up) and $\hsig^z_{20}$ (dashed: from canonical, crosses: from brute force calculation). Panels b) and c) both show the OSEE $S^{[m]}(t)$ between sites $m$ and $m+1$. Initial operators are b) $\hsig^z_{20}$ (calculated using the grand-canonical method) and c) $\hat{P}_{16}\hsig^z_{20}\hat{P}_{16}$ (calculated using the canonical method). The same data sets as in figure \ref{fig:delta08} are used.
}
\label{fig:cone}
\end{figure*}

As first example we take the spin-$\frac12$ XXZ chain
\begin{equation}
 \hH = -\frac12\sum_{\langle m,n\rangle} \left(\hsig^x_m\hsig^x_n+\hsig^y_m\hsig^y_n+\Delta\hsig^z_m\hsig^z_n\right),
\end{equation}
where the $\hsig$ denote the Pauli matrices and the sum runs over all nearest neighbours. The $\textrm{U}(1)$ symmetry (rotation around the z-axis) of the system implies conservation of the total magnetisation $\hM_z=\sum_m \hsig^z_m$. Via a Wigner-Jordan transformation this transforms into particle number conservation in the equivalent fermion lattice model.

The properties of the model depend strongly on the anisotropy $\Delta$. E. g., in the critical regime, $\vert \Delta \vert < 1$, spin transport is believed to be ballistic, while in the gapped regime it seems diffusive \cite{Prosen2009}. A quantity of interest in this context is the infinite temperature auto-correlation function (ITAC)
\begin{equation}
 \expv{\hOd_t\hO}_{T=\infty} = \Tr{\left[\hOd_t\hO\hat\rho_{T=\infty}\right]}
\end{equation}
for $\hO=\hsig^z$ at a given lattice site. The expectation value is taken at infinite temperature, which makes it straight forward to calculate it from the Heisenberg picture time evolution\footnote{Note that the infinite temperature density matrix of any system is proportional to the unity operator. Therefore from the definition of the ITAC, we see that in order to calculate it we have to do the same in the Schr\"odinger picture (take $\Tr{\left[\hOd \cdot \hat{U}_t\hO\mathbf{1}\hat{U}_t^\dagger\right]}$, where $\hat{U}_t$ is the full propagator) and in the Heisenberg picture (take $\Tr{\left[\hat{U}_t^\dagger\hOd\mathbf{1}\hat{U}_t \cdot \hO\right]}$). In fact this is an example where the requirement of using a mixed state in the Schr\"odinger picture makes it exactly as demanding as the Heisenberg picture calculation.}. In general it decays as $t^{-1/2}$, an observation usually called spin diffusion. In the spin-$\frac12$ chain this has been confirmed numerically for $\Delta \gtrsim 1$. Around $\Delta=1$ there is a change towards a $t^{-1}$ power law, which is the exact asymptotic behaviour at $\Delta=0$. However the asymptotics are hard to get numerically, especially for $\Delta$ around $1$ and larger, because of the limited timescales accessible. There exist exact diagonalization \cite{Fabricius1998}, as well as transfer matrix DMRG \cite{Sirker2006} studies. The Heisenberg picture t-DMRG results using unprojected operators presented here reproduce the results of the latter. The timescale accessible with Heisenberg picture t-DMRG is somewhat larger. For the value of $\Delta = 0.8$ we find an exponent for the decay of $\kappa\approx-0.83$ from the data shown in figure \ref{fig:delta08} using an empirical fitting function proposed in \cite{Fabricius1998},
\begin{equation}
 \expv{\hOd_t\hO}_{T=\infty} \approx t^\kappa\left[A+Be^{-\gamma(t-t_0)}\cos\left(\Omega(t-t_0)\right)\right],
\end{equation}
applied in a least squares fit to the data in the range $t=3$ to $t=11.5$. Although this value for the exponent is slightly closer to $-1$ than in previous calculations \cite{Fabricius1998, Sirker2006}, a decisive conclusion whether there is a sudden change of the exponent at $\Delta=1$ can not be drawn.

We calculate the ITAC here to compare the power of the different methods discussed above. In figure \ref{fig:delta08} we show Heisenberg picture t-DMRG results for the normalised ITAC at $\Delta=0.8$ both in the grand canonical ensemble,
\begin{equation}
 G(t)
 = \expv{\left(\hsig^z_m\right)_t\hsig^z_m}_{T=\infty}
 = \Tr{\left[\left(\hsig^z_m\right)_t\hsig^z_m\right]} / 2^L,
\end{equation}
as well as in the canonical ensemble,
\begin{equation}
 C_N(t) = \expv{\left(\hsig^z_m\right)_t\hsig^z}_{T=\infty} = \Tr{\left[\left(\hat{P}_N\hsig^z_m\hat{P}_N\right)_t\hsig^z_m\right]} \left/ \binom{L}{N} \right..
\end{equation}
Because $\Tr{\left[\left(\hat{P}_M\hO\hat{P}_N\right)_t\hO\right]}$ $=\Tr{\left[\hat{P}_M\hO_t\hat{P}_N\hO\right]}$ $=\Tr{\left[\hO_t\hat{P}_N\hO\hat{P}_M\right]}$ we can calculate the latter from both the projected, time-evolved or the unprojected, time-evolved $\hsig^z_m$. The behaviour in the canonical ensemble is as expected: For low filling, $C_N(t)$ decays only to a finite value. (From combinatorial arguments we find that $1-C_N(t) \le 4N/L$.) Therefore at half filling $C_N(t)$ has to be smaller than $G(t)$, because the latter is the weighted average
\begin{equation}
 G(t) = \frac{1}{2^L} \sum_{n=0}^N \binom{L}{n}C_n(t). 
\end{equation}

Figure \ref{fig:delta08} shows, that for low filling the canonical approach is clearly superior. However for generic filling ($N=L/2$ corresponds to a total magnetisation of 0) longer times can be reached with the grand-canonical algorithm. All curves shown are calculated using about the same computational resources. In order to propagate $\hat{P}_N\hsig^z_{20}\hat{P}_N$ for half filling up to the same point in time with the same accuracy as $\hsig^z_{20}$ an increase of computation time and memory by an order of magnitude would be required. The reason becomes apparent in figure \ref{fig:cone}a. The OSEE scales logarithmically both in the grand-canonical and the canonical picture for generic filling. This is expected from \cite{Prosen2007a, Muth2011}. In fact the OSEE looks the same for both $\hsig^z_{20}$ and $\hat{P}_N\hsig^z_{20}\hat{P}_N$, but the latter is shifted by the entanglement present in the initial MPO. The cut-off error\footnote{After each two-site operation of a Trotter step the evolved state $\ket{\Psi_j} = \hU_{m, m+1}(\Delta t)\ket{\Psi_{j-1}}$ has to be projected to the new reduced basis of dimension $\chi$. The resulting truncated state $\ket{{\rm RG}(\Psi_j)}$ (which is normalised before the next unitary is applied) has norm $\nu_j = \sqrt{\langle{\rm RG}(\Psi_j)\vert{\rm RG}(\Psi_j)\rangle}$ which fulfils $0\le1-\nu_j\ll1$. The accumulated cut-off error is defined as $1-\prod_j\nu_j$ which is approximately the sum of the single step cut-off errors, $1-\nu_j$, as long as it is much smaller than unity.\label{fn:cutoff}} in the algorithm therefore grows faster and the calculation breaks down earlier. In this example the higher bond dimension available for fixed particle number does not quite make up for this. Vice versa, to propagate $\hsig^z_{20}$ up to $t=20$, as can be done easily for $\hat{P}_N\hsig^z_{20}\hat{P}_N$ for low filling, would also require an increase of computational resources by orders of magnitude.

The OSEE as a function of both lattice position and time is shown in figure \ref{fig:cone}b-c. The light cone like appearance is imposed by causality\footnote{More rigorous result\cite{Bravyi2006, Eisert2006} of this reasoning have been provided in terms of Lieb-Robinson bounds.}. It confirms that there will be no finite size effects in the centre of the system before times close to 20. The projected operator is distinguished from the unprojected mainly by the fact, that there is initial entanglement away from the centre (which is where $\hsig^z_{20}$ acts nontrivial), compare to the illustration given in figure \ref{fig:entangled}. It is constant in time, as $\ket{\hat{P}_N}$ is an eigenstate of $\tilde{H}$. The entanglement generated dynamically seems to merely add.

Fig. \ref{fig:delta08} also shows a brute force calculation for $G(t)$, that does not take into account particle number conservation at all. It is clearly inferior to the unprojected method, section \ref{sec:ur}. Again a huge increase in computational resources would be required to reach the same accuracy.

\begin{figure*}[htb!]
\centering
\includegraphics[width=.7\textwidth]{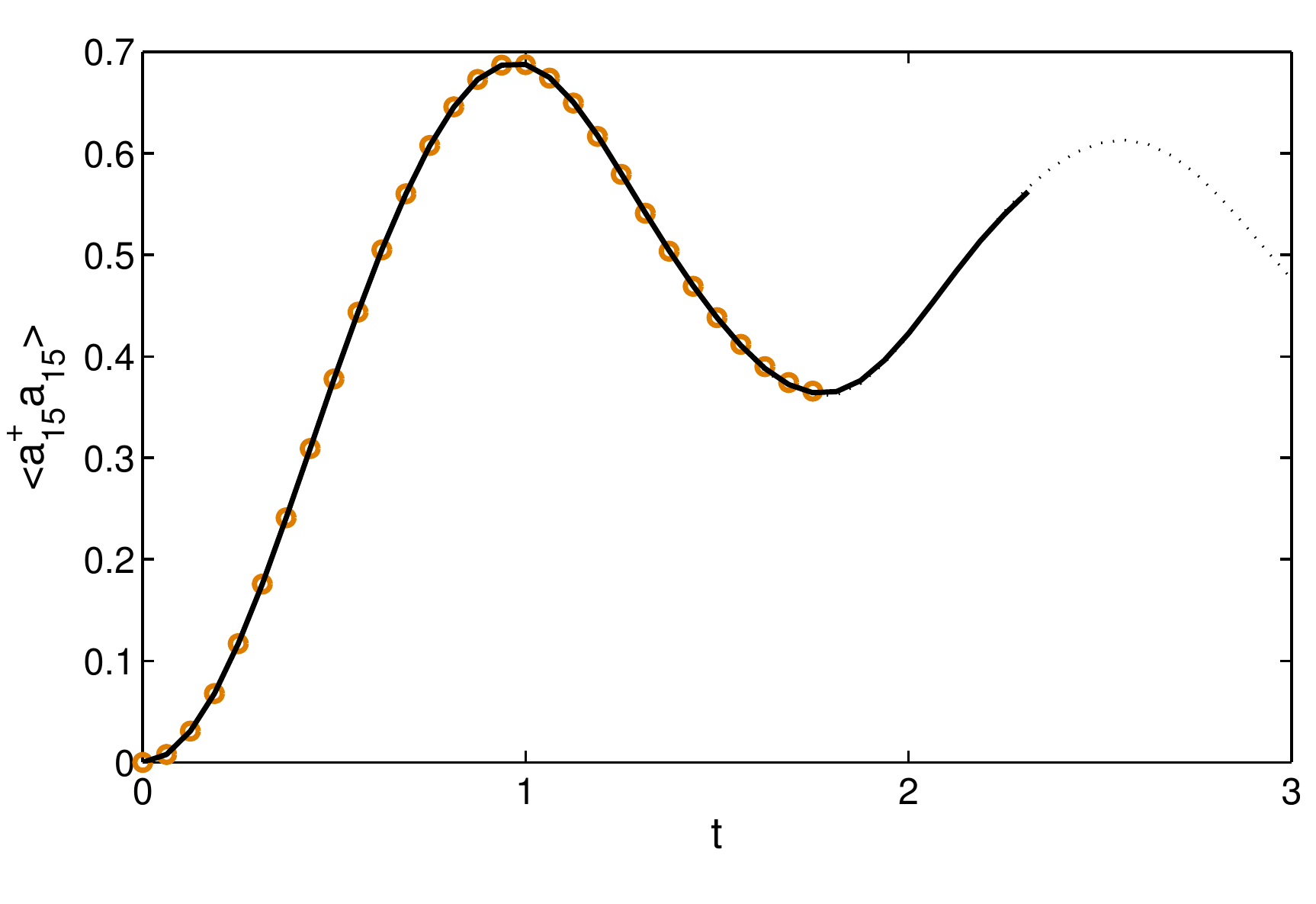}
\caption{
 Expectation value of the local density at site 15 on a Bose Hubbard lattice (restricted to local dimension $d=4$) of length $L=30$ at $U=10$ initially prepared in the state $\ket{0101\dots0101}$, calculated using the projected (solid, black) and the unprojected method (orange circles). Bond dimensions used where $\chi=2000$ in the canonical calculations, $\chi=500$ in the grand canonical. Both curves end at the point where the accumulated cut-off error (see footnote on page \pageref{fn:cutoff}) reaches $10^{-3}$. The dashed line shows the result of a Schr\"odinger picture calculation which can be regarded exact, as the cut-off error is numerically zero at this time scale. A TEBD \cite{Vidal2003} version of the t-DMRG algorithm is used with a fourth order trotter decomposition and time step size $1/16$ in all three cases.
}
\label{fig:bosons}
\end{figure*}

As an additional example we take the Bose Hubbard model,
\begin{equation}
 \hH = -J\sum_{\langle m,n\rangle}\left(\had_m \ha_{n}+h.a.\right) +\frac{U}{2}\sum_{m}\had_m \had_m \ha_{m} \ha_{m}.
\end{equation}
The $\had$ and $\ha$ operators are bosonic creation and annihilation operators. The first sum is again over nearest neighbours. For convenience, we set the hopping parameter $J=1$. Recently there is a lot of interest in the thermalization of far-from equilibrium states (not only in this model). Cramer et al. \cite{Cramer2008a} investigated the dynamics of the ``anti-ferromagnetic'' state $\ket{\Psi_0}=\ket{0101\dots0101}$ using Schr\"odinger picture t-DMRG. They propose an experimental setup to prepare this state and observe its dynamics in an experiment using ultracold atoms in optical lattices. First measurements have been reported recently \cite{Trotzky2011}.

Fig. \ref{fig:bosons} shows the dynamics of the local density at site $m=15$ in a system of total length $L=30$. Again the size has been chosen large enough, such that there are no boundary effects arriving at the centre for the times shown. The figure shows the expectation value in the state $\ket{\Psi_0}$ (where site 15 is empty) calculated using both the unprojected, $\bra{\Psi_0}\left(\had_{15}\ha_{15}\right)_t\ket{\Psi_0}$, and the projected method, $\bra{\Psi_0}\left(\hat{P}_{15}\had_{15}\ha_{15}\hat{P}_{15}\right)_t\ket{\Psi_0}$. Since $\ket{\Psi_0}$ is a particle number eigenstate, both expectation values are identical and coincide with a Schr\"odinger picture calculation, $\bra{\Psi_0}_t\had_{15}\ha_{15}\ket{\Psi_0}_t$.

Again both curves are calculated using approximately the same numerical resources. For performance purposes, we restrict the local Hilbert space to $d=4$. We find that using the projected operator we can calculate up to considerably larger times. So here the canonical method is ahead of the grand-canonical, \emph{even} if the particle number $N$ is of the order of $L/2$, in contrast to the first example. While the canonical method is only moderately affected by a higher local dimension $d>4$ ($d=16$ being the largest meaningful number here), the unprojected one breaks down as the Hilbert space dimension increases (not shown in the figure).

Although the timescales reachable are not large enough to see the local density equilibrate at $\frac12$, a Schr\"odinger picture calculation is actually the method of choice in this example, as the timescale reachable is still significantly larger\cite{Cramer2008a} than in the Heisenberg picture. This is true in spite of the fact, that the local density is a conserved density and in the Heisenberg picture we therefor expect much better scaling of the OSEE with time\cite{Muth2011}. It can be explained by the overhead of having to include high local occupation numbers in the Heisenberg picture (in contrast to the above spin-$\frac12$ example), which are actually not populated dynamically for the given initial state. This is a quite general drawback of the Heisenberg picture calculation whenever the local degree of freedom is large, not necessarily because the particles are bosons, but also, e.g., for higher spin models. In the case of a highly entangled or mixed initial state, the two pictures might compare differently. Another overhead introduced by allowing for higher local occupation numbers is the introduction of higher energy scales, because the maximum local interaction energy in the truncated Bose Hubbard model is $\frac{U}{2}(d-1)(d-2)$. Therefor time steps $\Delta t$ must be reduced as $d^{-2}$ if a Suzuki Trotter expansion is used, to keep track of the time evolution correctly.

\section{Conclusion}
\label{sec:conc}

The two different approaches to include particle number conservation into an MPO have quite different effects on the performance of a Heisenberg-picture t-DMRG. The grand-canonical method brings the advantages known from ordinary (t-)DMRG, namely, the reduction of the Hilbert space dimension without introducing any additional entanglement. It is the method of choice in the presence of an appropriate symmetry. The Hilbert space dimension can be further reduced by projecting the MPO to a certain symmetry-sector. The effect then is quite counter-intuitive. Although the projected operators do only contain a small subset of the information present in the grand-canonical MPO, their propagation in time is not always easier. This is due to the entanglement introduced by fixing the total particle number. (The identity is not a product operator if projected to a symmetry sector.) In the low filling case, the reduction of the Hilbert space dimension is more important and we gain access to longer times. For generic filling however, the grand-canonical method remains superior.

\section*{Acknowledgements}

I am indebted to M. Fleischhauer for valuable discussions during the completion of this work.


\section*{References}



\providecommand{\newblock}{}

\end{document}